\documentclass[final,3p,times,twocolumn,longtitle]{elsarticle}

\journal{Nuclear Instrument and Method}

\usepackage{graphicx}
\usepackage{multicol}
\usepackage{multirow}
\usepackage{lineno}

\footnotetext[1]{}

\begin{document}

\begin{frontmatter}

\title{Study on the large area MCP-PMT glass radioactivity reduction }

\author[IHEP,UCAS]{Xuantong Zhang}
\author[IHEP]{Jie Zhao\corref{mycorrespondingauthor}}
\cortext[mycorrespondingauthor]{Corresponding author}
\ead{zhaojie@ihep.ac.cn}
\author[IHEP]{Shulin Liu}
\author[IHEP]{Shunli Niu}
\author[NNVT]{Xiaoming Han}
\author[IHEP]{Liangjian Wen}
\author[IHEP,NCEPU]{Jincheng He}
\author[IHEP]{Tao Hu}

\address[IHEP]{Institute~of~High~Energy~Physics, Beijing 100049, China}
\address[UCAS]{University of Chinese Academy of Sciences, Beijing 100049, China}
\address[NNVT]{North Night Vision Technology Co., Ltd., Nanjing 211106, China}
\address[NCEPU]{North China Electric Power University, Beijing 102206, China}

\begin{abstract}
The Jiangmen Underground Neutrino Observatory (JUNO) will install about 18,000 20-inch Photomultiplier Tubes (PMTs) in the center detector to achieve 3\%/$\sqrt{E(MeV)}$ energy resolution. From the full detector Monte Carlo (MC) simulation, besides the liquid scitillator (LS) and Acrylic nodes, PMT glass has the largest contribution to the natural radioactive background. Various technologies have been developed in the Chinese industry to control the environment and to improve the production process. We have monitored the glass production for more than two months, and the radioactivity in glass was measured using a low background gamma ray spectrometer equipped with a high resolution HPGe detector. The $^{238}$U, $^{232}$Th and $^{40}$K of the glass bulb are reduced by a factor of 2, 9 and 15 respectively, and now they can reach 2.5 Bq/kg for $^{238}$U, 0.5 Bq/kg for $^{232}$Th and 0.5 Bq/kg for $^{40}$K.
\end{abstract}


\end{frontmatter}
\section{Introduction}
\par
In order to detect weak signals, PMT is widely used in high energy physics experiments. In large neutrino experiment, such as SuperK, KamLAND, Borexino, JUNO, Dayabay, the PMT with large diameter is one of the key components. Development of the large PMT is extremely difficult due to the complex production technology. In the past, the high-performance 20-inch PMT can only be produced by Hamamatsu company in Japan.

JUNO experiment aims to determine the neutrino mass hierarchy by detecting the oscillation of antineutrinos from the reactors. About 18,000 20-inch PMTs will be installed on the center detector to achieve 3\%/$\sqrt{E(MeV)}$ energy resolution, and 2,000 in the water pool to record the cosmic muons. The Chinese PMT group has now developed the 20-inch PMT with Micro Channel Plates (MCP), produced by NNVT company~\cite{Qian:mcp-pmt}. JUNO plans to use about 5,000 Hamamatsu PMTs and 15,000 NNVT MCP-PMTs.

The neutrino experiment requires low background. Radioactivity from all used materials will form isolated signals (singles) in the target and mimic neutrino signals. Better control and shielding of the radioactivity from materials is important in the neutrino experiment. The comparison of the radioactivity level in the PMT between Hamamatsu and NNVT is listed in Table~\ref{tab:bkgBudget}, and the radioactivity of the NNVT PMTs is based on our latest experimental result. The NNVT company will produce the 20-inch PMT with the lowest radioactivity in the world.

\section{JUNO Experiment Requirement}
\par
The principle of the antineutrino detection is the inverse beta decay reaction, and the feature is prompt-delayed signals correlated both in time, space and energy. After the event selection, about 60 antineutrinos can be detected by the JUNO detector per day~\cite{An:2015jdp}. Extremely better control on the background is needed for JUNO. Two isolated signals can accidentally fall into the correlated time window and mimic the antineutrino events, which is called the accidental background. The singles mainly come from the natural radioactivity and long-lived cosmogenic isotopes. In order to reach our physical goal, we need to control the singles rate in the fiducial LS target to be less than 10 counts per second (cps), which will lead to about one accidental background per day.

JUNO will use various kinds of materials for the central detector and the veto detector. Different materials have different levels of natural radioactivities, such as $^{238}$U, $^{232}$Th, $^{40}$K. The full detector simulation for the main materials considering the possible radioactive levels is done in~\cite{MC:2016cpc}. In order to reduce the radioactive background, our fiducial volume for the analysis is set to be R $<$ 17.2 m and energy cut of larger than 0.7 MeV. Table~\ref{tab:bkgBudget} shows the comparsion of the radioactivity and singles rate between Hamamatsu PMT and NNVT PMT, assuming one quarter of the 20-inch PMTs from Hamamatsu and the rest from NNVT.

\begin{table}[htb]
\begin{center}
\footnotesize
\tabcolsep2.1pt
\renewcommand\arraystretch{1.3}
    \begin{tabular}{c|c|ccc|c}
            \hline
			Material & Mass & $^{238}$U & $^{232}$Th & $^{40}$K & Singles in FV(cps)\\
            \hline	
            Hamamatsu &33t& 400ppb & 400ppb & 40ppb & 0.68 \\
			NNVT & 100t & 202ppb & 123ppb & 3.54ppb & 0.79\\
          \hline
    \end{tabular}
    \caption{Comparison of the radioactivity between Hamamatsu PMT and NNVT PMT, and the values are our bidding requirements. The singles rate in the central detector with fiducial cut are scaled from~\cite{MC:2016cpc}.}
	\label{tab:bkgBudget}
\end{center}
\end{table}

The mass of one NNVT PMT is about 7.5 kg. Natural radioactivity from PMTs need to be carefully controlled due to its large usage. The impact on the sensitivity ($\Delta\chi^{2}$) of the mass hierarchy determination from the radioactivity of the PMT glass will be increased by about 0.1 when we use only one quarter of Hamamatsu PMTs.

\section{Experiment on the Improvement of Glass Bulb Production}
\par
\begin{figure}[htb]
\begin{center}
	\includegraphics[width=6cm]{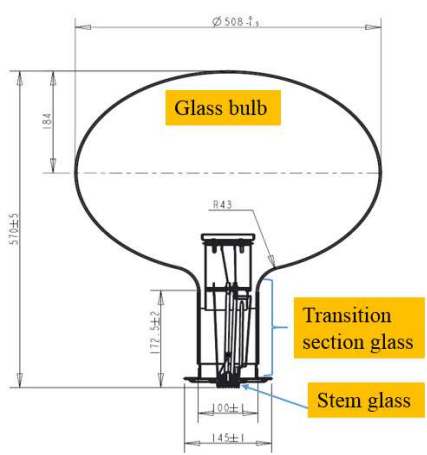}
	\caption{The diagram of the PMT tube.}\label{fig:glass}
\end{center}
\end{figure}

Three components of the NNVT PMT use glass, including blub glass, transition section glass and stem glass, as shown in Figure~\ref{fig:glass}. The mass of the glass is larger than 92\% of the whole PMT mass. The mass percentage of each part of the MCP-PMT is shown in Table~\ref{tab:masspercentage]}.

\begin{table*}[htb]
\begin{center}
\footnotesize
	\begin{tabular}{c|c|c|ccc|ccc}
        \hline
		No. & Composition & Percentage & $^{238}$U & $^{232}$Th & $^{40}$K & \multicolumn{3}{c}{Expected contribution(\%)}\\
        &&of Mass(\%) & (Bq/kg) & (Bq/kg) & (Bq/kg) & $^{238}$U & $^{232}$Th & $^{40}$K\\
		\hline
		1 & Borosilicate Glass Bulb &  89.48 & 2.5 & 0.5 & 0.5 & 90 & 90 & 45\\
		2 & Transition Section &  2.82 & 5 & 2 & 15 & 5.6 & 11 & 42\\
		3 & 95\% ceramics &  0.52 & 10.6 & 4.2 & 0.46 & 2.2 & 4.4 & 0.2\\
		4 & 99\% ceramics &  0.46 & 2.7 & 1.3 & 1.96 & 0.5 & 1.2 & 0.9\\
		5 & Stem Glass &  0.33 & 5 & 2 & 15 & 0.7 & 1.3 & 5.0 \\
		6 & MCP &  0.02 & - & - &216.7 & && 4.3\\
		7 & Other &  6.36 & - & - & - &  &&\\
		\hline
		& Total &  100 & 2.5 & 0.5 & 1 & &&\\ \hline
	\end{tabular}
	\caption{Mass percentage and radioactivity of the main MCP-PMT components. Glass components (Glass Bulb, transition section and stem) have $>$92\% mass proportion in the whole PMT. Background reduction of the glass can efficiently reduce the whole background of the PMT.}
	\label{tab:masspercentage]}
\end{center}
\end{table*}

\subsection{Bulb Glass}
\par
\begin{figure}[htb]
\begin{center}
	\includegraphics[width=7cm]{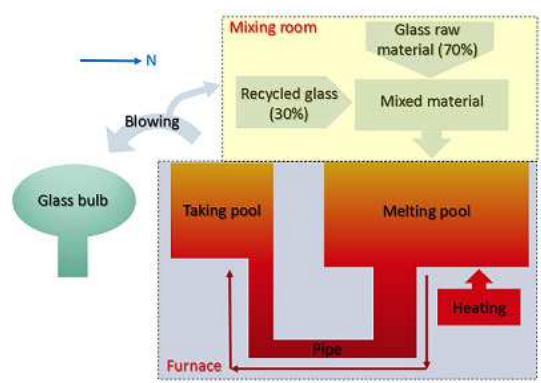}
	\caption{The diagram of the furnace.}\label{fig:furnace}
\end{center}
\end{figure}

The glass bulb is produced in a glass furnace, as shown in Figure~\ref{fig:furnace}. The furnace is built as an U-style connected vessel using fused zirconia alumina brick. The north side of the furnace is the melting tank, while the south side is the taking tank. The melting tank is heated by electric heating wire and the temperature can reach around 1,500 degree Celsius. The taking tank has three taking holes and only the east hole is used to produce the glass bulb.

The components of the raw material are listed in Talbe~\ref{tab:glassComp}. The raw material is mixed with about 30\% of the recycled glass for the production of the glass bulb. The whole furnace can melt 4.8 tons raw material everyday. The volume of the furnace can hold 15 tons of raw material, which means the raw material will stay in the furnace for 3 days until it is taken from the taking tank. All the raw materials are low background materials, as shown in Table~\ref{tab:quartz}. However, the furnace brick is made by the zirconia alumina brick with very high radioactivity. During the production, part of the brick may be dissolved into the liquid glass and introduce the background isotopes.

\begin{table}[htb]
\begin{center}
\footnotesize
	\begin{tabular}{c|c|c}
\hline
    Composition & Weight(\%) & Introduced compounds \\ \hline
    SiO$_{2}$ & $\sim$80 & Quartz sand \\
    B$_{2}$O$_{3}$ & $\sim$13 & Borax and Boric-acid \\
    Na$_{2}$O & $\sim$4 & Borax \\
    Al$_{2}$O$_{3}$ & $\sim$3 & Aluminum hydroxide \\
    NaCl & $\sim$0.1 & Industrial salt \\ \hline
	\end{tabular}
	\caption{The components of the raw material for the production of glass bulb.}
	\label{tab:glassComp}
\end{center}
\end{table}

\begin{table}[htb]
\begin{center}
\footnotesize
	\begin{tabular}{c|ccc}
\hline
		Bq/kg & $^{238}$U  & $^{232}$Th  & $^{40}$K  \\
		\hline
		Fused alumina  & \multirow{2}{*}{1880$\pm$30} & \multirow{2}{*}{322$\pm$11} & \multirow{2}{*}{Covered} \\
        zirconia brick &&& \\ \hline
		Quartz sand & $<$0.06 & $<$0.05 & 0.05$\pm$0.1 \\ \hline
		Raw material  & $<$0.06 & $<$0.05 & 0.32$\pm$0.1 \\ \hline
	\end{tabular}
	\caption{The radioactivity of fused alumina zirconia brick and raw materials. The peak of $^{40}$K in the fused alumina zirconia brick is covered by other peaks in the U/Th chains.}
	\label{tab:quartz}
\end{center}
\end{table}

Before this experiment, the radioactivity of the glass bulb is up to 5.4 Bq/kg for $^{238}$U, 4.3 Bq/kg for $^{232}$Th and 7.6 Bq/kg for $^{40}$K. Since the raw materials are all low background, the background isotopes in the final glass products mainly come from the production procedure. Figure~\ref{fig:furnace} shows the procedure of the glass bulb production. Most of the $^{238}$U and $^{232}$Th are attributed to the dust, concrete and zirconia alumina brick of the furnace. It is not easy to replace the furnace brick with low radioactivity materials, so we can only try to control dust and concrete to reduce the radioactivity background and regularly discharge the glass from the bottom of the furnace to reduce the radioactivity accumulated in the furnace. The $^{40}$K background mostly came from the raw material recipe and the cooling water, so we exchanged the underground water by the deionized water in order to improve the cooling system.

In general, four improvements are applied to reduce the radioactivity of the bulb glass.

\begin{itemize}
	\item Raw materials with low radioactivity;
	\item Use steel plate and plastic cover to isolate the raw material from the concrete and dust, which contain large amount of Uranium and Thorium;
	\item Cooling the glass with deionized water to get rid of the potassium impurity;
	\item Discharge the liquid glass from the bottom of glass-taking pool every week to avoid the radioactivity isotopes accumulation.
\end{itemize}

\subsection{Transition Section Glass and Stem Glass}
\par
Due to the different thermal expansion ability of the glass and kovar metal, transition section is designed to avoid broken in the glass-kovar connector.

The transition section consists of four types of glass with different thermal expansion coefficients. The coefficients are adapted by the components of glass, especially the potassium component. Table~\ref{tab:allresults} shows the components and radioactivity in different transition sections and stem glass.

\begin{table}[htb]
\begin{center}
\footnotesize
	\begin{tabular}{cc|ccc}
        \hline
		\multicolumn{2}{c|}{Bq/kg} & $^{238}$U  & $^{232}$Th  & $^{40}$K  \\
		\hline
		\multirow{4}{*}{TranSec} & H1 & 0.74$\pm$0.16 & 0.29$\pm$0.02 & 9.61$\pm$0.60 \\
		& H2 & 1.50$\pm$0.22 & 0.75$\pm$0.03 & 4.73$\pm$0.37 \\
		& H3 & 4.24$\pm$0.38 & 0.96$\pm$0.04 & 7.50$\pm$0.50 \\
		& H8 & 0.95$\pm$0.16 & 0.54$\pm$0.03 & 2.45$\pm$0.25 \\
		\hline
		\multicolumn{2}{c|}{Stem glass} &0.78$\pm$0.15 & 0.34$\pm$0.02 & 3.80$\pm$0.31 \\
		\hline
		\multicolumn{2}{c|}{Requirement}& $<$5 & $<$2 & $<$15 \\
        		\hline
	\end{tabular}
	\caption{Radioactivity in different transition sections and stem glass}
	\label{tab:allresults}
\end{center}
\end{table}

The background of transition sections and stem contribute about half potassium background to the whole PMT glass.

\section{Radioactivity Measurement System}
\par
In order to make sure the measurement is exactly matched to the real bulb product, most of the glass samples are directly taken from the furnace during the production of the glass bulb. The dimensions of the sample is 120mm in diameter and 50mm in height, which can match our $\gamma$ spectrometer.

\subsection{High-Purity Germanium Detector}
\par
High-Purity Germanium(HPGe) detector is a high-precision $\gamma$ spectrometer used to analyze the radioactivity of samples. Our spectrometer is a four inch well-type HPGe detector from Canberra~\cite{Niu:2014cpc}. The resolution is 2.1 keV at 1.33 MeV. The nominal size of the Ge crystal is of 86.3 mm diameter and 85.8 mm height with a 4.89 mm front gap, and the aluminum shell is 1.5 mm thickness. Surrounding the Ge crystal, four layers of walls are made for anti-Compton and shielding the $\gamma$ background. They are BGO crystal, oxygen-free copper, lead and plastic scintillator from inner to exterior. Geant4 simulation is used to correct the analysis efficiency in the HPGe detector system.

$^{238}$U and $^{232}$Th are measured by their daughter isotopes in their decay chains, assuming the decay chains are at equilibrium. For $^{238}$U, three daughter isotopes, $^{226}$Ra, $^{214}$Pb and $^{210}$Bi, are measured. For $^{232}$Th, also three daughter isotopes, $^{228}$Ac, $^{212}$Pb and $^{208}$Tl, are measured.

\subsection{Radon Decay in Glass}
\par
\begin{figure}[htb]
\begin{center}
	\includegraphics[width=7.5cm]{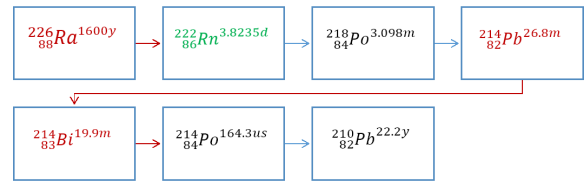}
	\caption{Part of $^{238}$U decay chain. The $^{222}$Rn has short half life. If Radon releases from glass, the equilibrium state will be broken and need time to be regained.}\label{fig:udecay}
\end{center}
\end{figure}

 The samples must be placed for at least 1000 hours to reach equilibrium state for the $^{238}$U decay chain. The non-equilibrium state is caused by the radon releasing. When the liquid glass is in the furnace, the radon releases from the liquid glass because of low solubility at high temperature. As shown in Figure~\ref{fig:udecay}, $^{222}$Rn with half life of 3.82 days is the daughter isotope of $^{226}$Ra with 1600 years half life, and all the daughter isotopes of $^{222}$Rn have short half lives except $^{210}$Pb. Because of the short half lives, those daughter isotopes of $^{222}$Rn will quickly decay when the $^{222}$Rn releases from liquid glass, which leads to the non-equilibrium state of $^{238}$U decay chain. After the glass modeling, radon releasing from glass is stopped and the regain of equilibrium state will last at least 1000 hours, which depends on the remaining proportion of Radon in glass and the half life of $^{222}$Rn.

However, in order to return the measurement results timely and save the cost of glass production, we don't have enough time to keep the samples placed for that long time. As we have already known the half lives of the daughter isotopes in $^{238}$U decay chain, we can calculate radioactivity as a function of time. Assuming the remaining proportion of radon is $m$, a percentage from 0 to 100\%, the radioactivity of $^{226}$Ra and $^{222}$Rn are $$A_{^{226}Ra}(t)=\lambda_{Ra}N_0e^{-\lambda_{Ra}t},$$ $$A_{^{222}Rn}(t)=\frac{\lambda_{Ra}N_0}{\lambda_{Rn}-\lambda_{Ra}}(\lambda_{Rn}e^{-\lambda_{Ra}t}-(\lambda_{Ra}-m(\lambda_{Rn}-\lambda_{Ra}))e^{-\lambda_{Rn}t}),$$ where $A_i$ is the radioactivity of $i$-isotope, $t$ is the time, $\lambda_i$ is the decay constant of $i$-isotope and $N_0$ is the initial atoms of $^{226}Ra$. Considering that $\lambda_{Rn}$ is largely greater than $\lambda_{Ra}$, the radioactivity ratio between $^{222}$Rn and $^{226}$Ra can be approached by $$Ratio=\frac{A_{^{222}Rn}(t)}{A_{^{226}Ra}(t)}\approx1-(1-m)e^{-\lambda_{Rn}t}.$$

\begin{figure}[htb]
	\centering
	\includegraphics[width=8cm]{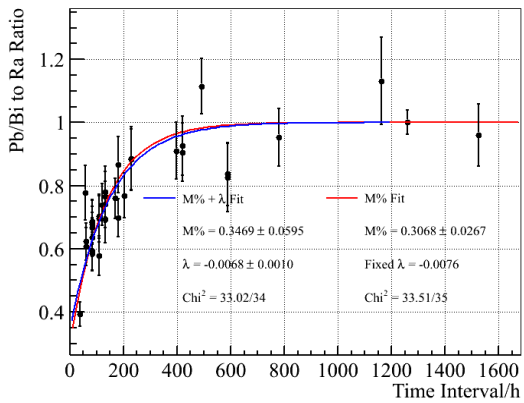}
	\caption{The ratio of $^{214}$Bi/$^{214}$Pb to $^{222}$Rn as a function of the time interval between taking and measuring of each sample. Two fitting methods are used. The red curve fixes the half life of $^{222}$Rn at real value and fits remaining proportion of radon. The blue curve fits both the remaining proportion and half life of $^{222}$Rn.}\label{fig:fitting}
\end{figure}

We have measured three isotopes in the U-chain, two isotopes after $^{222}$Rn and one isotope before. Figure~\ref{fig:fitting} shows the ratio of radioactivity as a function of the time interval between taking and measuring of each sample. With the known decay constant of $^{222}$Rn, the $\lambda_{Rn}$ can be either fixed in the true value or released to fit, then the remaining proportion of radon $m$ is fitted. Table~\ref{tab:fitresult} shows the fitting results of the remaining proportion and decay constant of $^{222}$Rn. The two fitting curves are close and the fitting results have only 1$\sigma$ difference.

\begin{table}[htb]
\begin{center}
\footnotesize
	\begin{tabular}{c|ccc}
        \hline
		Method & $\lambda_{Rn}$ & $m$ & $\chi^2/ndf$ \\
        \hline
		$m$ Only Fit & 0.0076(Fixed) & 0.307$\pm$0.027 & 33.51/35 \\

		$m+\lambda_{Rn}$ Fit & 0.0068$\pm$0.0010 & 0.347$\pm$0.060 & 33.02/34 \\ \hline
	\end{tabular}
	\caption{The fitting results of radon decay. $m$ only fit and $m+\lambda_{Rn}$ results correspond to red curve and blue curve respectively in Figure~\ref{fig:fitting}. The results of the two methods are consistent with each other in 1$\sigma$.}
	\label{tab:fitresult}
\end{center}
\end{table}

\subsection{Radon Re-measurement}
\par
In order to further prove that radon is at non-equilibrium state at the begining, we remeasured two samples with time interval larger than 1000 hours. Table~\ref{tab:radonremeasurement} shows the remeasured results compared with their initial results.

\begin{table}[htb]
\begin{center}
\footnotesize
	\begin{tabular}{cccc}
        \hline
		Sample Date & Measurement Date & $^{214}$Bi/$^{214}$Pb & $^{226}$Ra \\
		\hline
		\multirow{2}{*}{Dec.05.2016} & Dec.08.2016 & 1.37$\pm$0.05 & 2.1$\pm$0.2 \\
		& Mar.14.2017 & 2.0$\pm$0.1 & - \\
		\hline
		\multirow{2}{*}{Dec.19.2016} & Dec.20.2016 & 1.40$\pm$0.04 & 2.0$\pm$0.2 \\
		& Mar.15.2017 & 2.0$\pm$0.1 & - \\ \hline
	\end{tabular}
	\caption{Radon decay remeasured results. In order to make sure the decay chain was at non-equilibrium state when the samples were measured at the first time, re-measurement was carried out after the first measurement for about three months, which is enough to regain the equilibrium state.}
	\label{tab:radonremeasurement}
\end{center}
\end{table}

 Since the time interval between taking the sample and the re-measurement is longer than 1000 hours, the equilibrium state has been achieved inside the samples and the radioactivity of $^{222}$Rn equals to $^{238}$U in the samples. The radioactivity of $^{214}$Bi/$^{214}$Pb in the re-measurement is consistent with the initial result of $^{226}$Ra. For all the samples, $^{226}$Ra is used as the measured isotope due to the equilibrium between $^{238}$U and $^{226}$Ra.
\subsection{Radioactivity Measurement Summay}
\par
As what we mentioned above, there are two methods to measure the $^{238}$U radioactivity. Both of them have advantages and disadvantages.

For $^{214}$Pb/$^{214}$Bi method, the uncertainty of radioactivity is smaller due to the high statistics and clear peaks of $^{214}$Pb and $^{214}$Bi decay events. However, if the decay chain is not at the equilibrium state, the $^{214}$Pb/$^{214}$Bi method can not give precise center value due to the inaccurate time interval.

For $^{226}$Ra method, the gamma peak of $^{226}$Ra at 186 keV has overlap with $^{235}$U gamma peak at 185 keV. The HPGe detector can not distinguish those two peaks, because the resolution of the HPGe detector is 2 keV. To solve this problem, we assumed that the abundance of $^{238}$U and $^{235}$U in glass are same with their natural abundance, then we can correct the gamma peak and measure the $^{238}$U. As a result, the uncertainty will be larger with lower statistics and overlap.

To measure the $^{238}$U precisely, we use both $^{226}$Ra and $^{214}$Pb/$^{214}$Bi method in our measurement. Because of the high precision, $^{214}$Pb/$^{214}$Bi method is used when the glass is at equilibrium state. Otherwise, $^{226}$Ra method is used to get rid of the non-equilibrium effect.

\section{Experiment Results}
\par
\begin{figure}[htb]
\begin{center}
	\includegraphics[width=\linewidth]{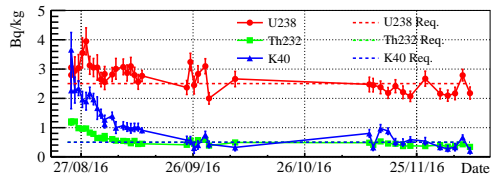}
	\caption{Background results from Aug.~2016. The experiment was stopped in October, and the company started the mass production of the glass bulb in December. The requirement is decided based on the whole detector background Monte Carlo simulation results and experiment results. $^{238}$U is stable at 2.5 Bq/kg, and $^{232}$Th and $^{40}$K are less than 0.5 Bq/kg.}\label{fig:uthkresults}
\end{center}
\end{figure}

The radioactivity reduction experiment started from Aug. 2016 at the glass factory in China. The control was strictly carried out and samples taken from the furnace were sent to IHEP for radioactivity measurement every day. The results are shown in Figure~\ref{fig:uthkresults}. Both $^{232}$Th and $^{40}$K have large improvement in about one month. However, the improvement of $^{238}$U was very limited, and we suspect this may due to the high radioactivity from the fused alumina zirconia brick in the furnace during the melting.

\section{Summary}
\par
Large amount of 20-inch PMTs will be installed in the JUNO detector. In order to achieve the PMTs with low radioactivity, we have done many efforts on the control of the glass bulb production, including the environment and the production process. Since the $^{238}$U decay chain is not at equilibrium state during the measurement for many samples, we have developed an improved method for the data analysis to achieve the accurate results. The $^{214}$Pb/$^{214}$Bi peaks are used when the glass is at equilibrium state. Otherwise, $^{226}$Ra method is used to get rid of the non-equilibrium effect. With a pilot experiment lasted for about two months experiment, the $^{238}$U, $^{232}$Th and $^{40}$K of the glass bulb are reduced by a factor of 2, 9 and 15 respectively, and now they can reach 2.5 Bq/kg for $^{238}$U, 0.5 Bq/kg for $^{232}$Th and 0.5 Bq/kg for $^{40}$K, which are several times lower than 20-inch Hamamatsu PMTs.

From the beginning of 2017, the company started the mass production of PMT bulbs. We continued to monitor the radioactivity of the glass bulb with a frequency of one sample per week for the quality control. By now, the radioactivity is stable at our expected values.

\section{Acknowledgement}
\par
This work is supported by the the Strategic Priority Research Program of the Chinese Academy of Sciences, Grant No. XDA10010400, Postdoctoral Science Foundation of China and Chinese Academy of Sciences (2015IHEPBSH101). The authors would like to acknowledge Dr. Xiao Cai for his technical support, Dr. Baojun Yan for his help in the earlier stage.

\section*{References}
\par

\end{document}